\documentclass[
showpacs,
floatfix,
aps,
prb,
amsmath,
twocolumn,
tightenlines
groupaddress,
]{revtex4}
\usepackage{amsmath,amssymb}
\usepackage{amscd, latexsym}
\usepackage{mathrsfs}
\usepackage{graphicx} 
\usepackage{epstopdf} 
\usepackage{cancel} 
\usepackage{amsfonts}
\usepackage{exscale}
\usepackage{dcolumn}
\usepackage{bm}
\usepackage{color}

\newcommand{\ket}[1]{\left| #1 \right>} 
\newcommand{\bra}[1]{\left< #1 \right|} 

\newcommand{\be}{\begin{equation}}
\newcommand{\ee}{\end{equation}}
\newcommand{\bea}{\begin{eqnarray}}
\newcommand{\eea}{\end{eqnarray}}

\DeclareMathOperator{\Tr}{Tr}

\begin{document}

\title{Quantum efficiency of a microwave photon detector based on a current-biased Josephson junction}

\author{A. Poudel}
\author{R. McDermott}
\author{M. G. Vavilov}
\affiliation{Department of Physics, University of Wisconsin-Madison, Wisconsin 53706, USA }

\date{\today}

\begin{abstract}
We analyze the quantum efficiency of a microwave photon detector based on a current-biased Josephson junction. We consider the Jaynes-Cummings Hamiltonian to describe coupling between the photon field and the junction. We then take into account coupling of the junction and the resonator to the environment. We solve the equation of motion of the density matrix of the resonator-junction system to compute the quantum efficiency of the detector as a function of detection time, bias current, and energy relaxation time. Our results indicate that junctions with modest coherence properties can provide efficient detection of single microwave photons, with quantum efficiency in excess of 80\%.

\end{abstract}


\maketitle

\section{Introduction}
Quantum optical photodetectors have contributed significantly to the development of quantum optics and atomic physics~\cite{WallsBook} and now play an essential role in optical quantum information applications such as quantum computing and quantum key distribution~\cite{Knill01, Kok07, Hadfield09}. Recently, circuit quantum electrodynamics (cQED) has emerged as a novel paradigm for the study of radiation-matter interaction in mesoscopic systems~\cite{Blais04, Wallraff04, You11}. Moreover, cQED is an attractive candidate for scalable quantum computing and transmission of quantum information~\cite{Majer07, You03, Buluta11}. Following the original proposal, a variety of cQED architectures demonstrating strong coupling between single photons and superconducting integrated circuits have been realized experimentally~\cite{Houck07, Chiorescu04}. This work has paved the way for the development of  a superconducting microwave photon detector~\cite{Romero09, Chen11, Peropadre11} with possible applications to quantum information processing and communication~\cite{Gisin02}.

The microwave photon detector is based on the current-biased Josephson junction (JJ): the JJ is biased so that absorption of a single microwave photon induces a transition to the voltage state, producing a large and easily detected classical signal~\cite{Chen11}. While these detectors are straightforward to operate and show potential for scalability, performance is degraded by spurious dark counts due to quantum tunneling events in the absence of an absorbed photon; moreover, energy relaxation within the detector results in photon loss and leads to a reduction in measurement efficiency.
\begin{figure}
\includegraphics[width=0.42\columnwidth]{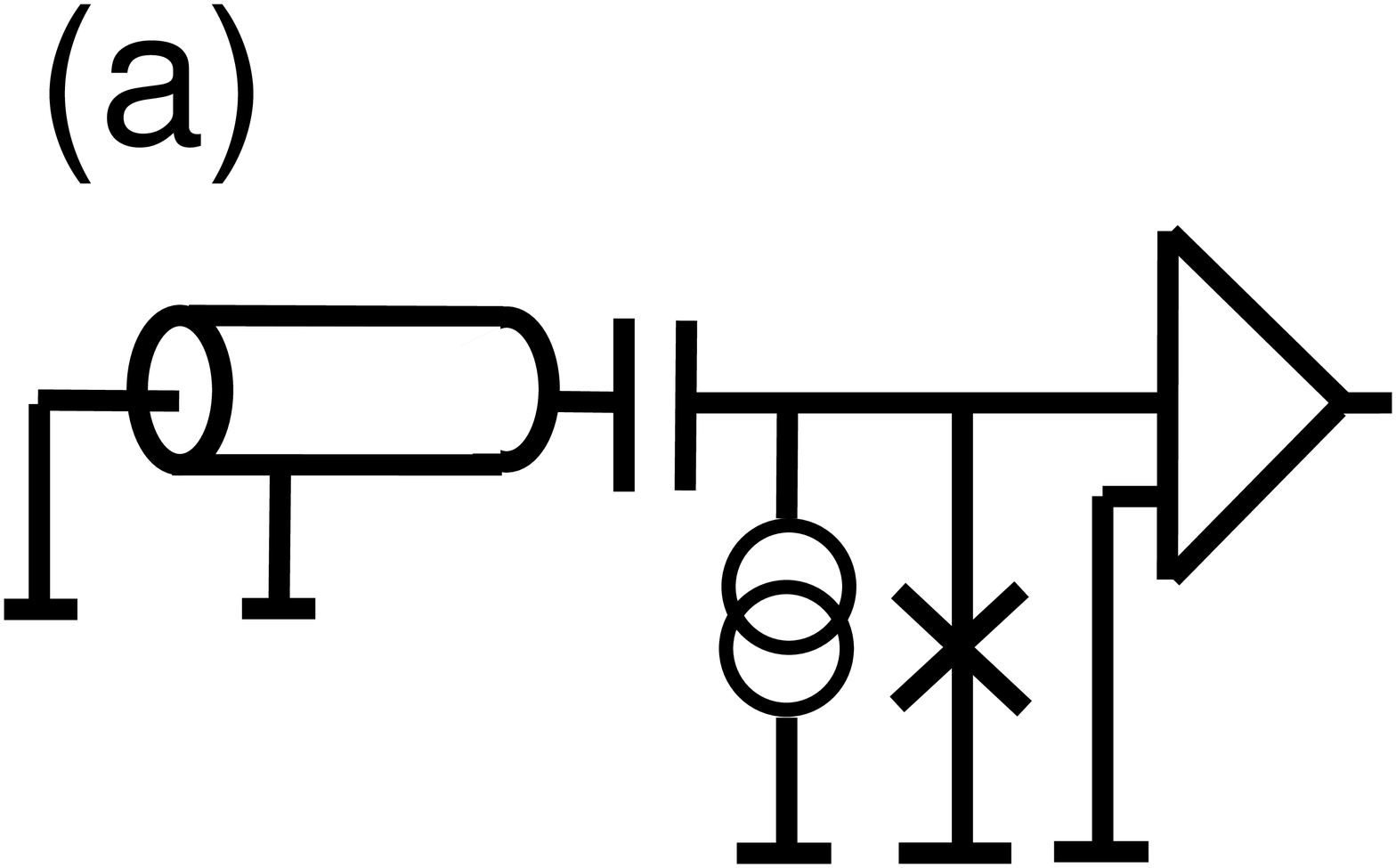}\hspace{5mm}
\includegraphics[width=0.42\columnwidth]{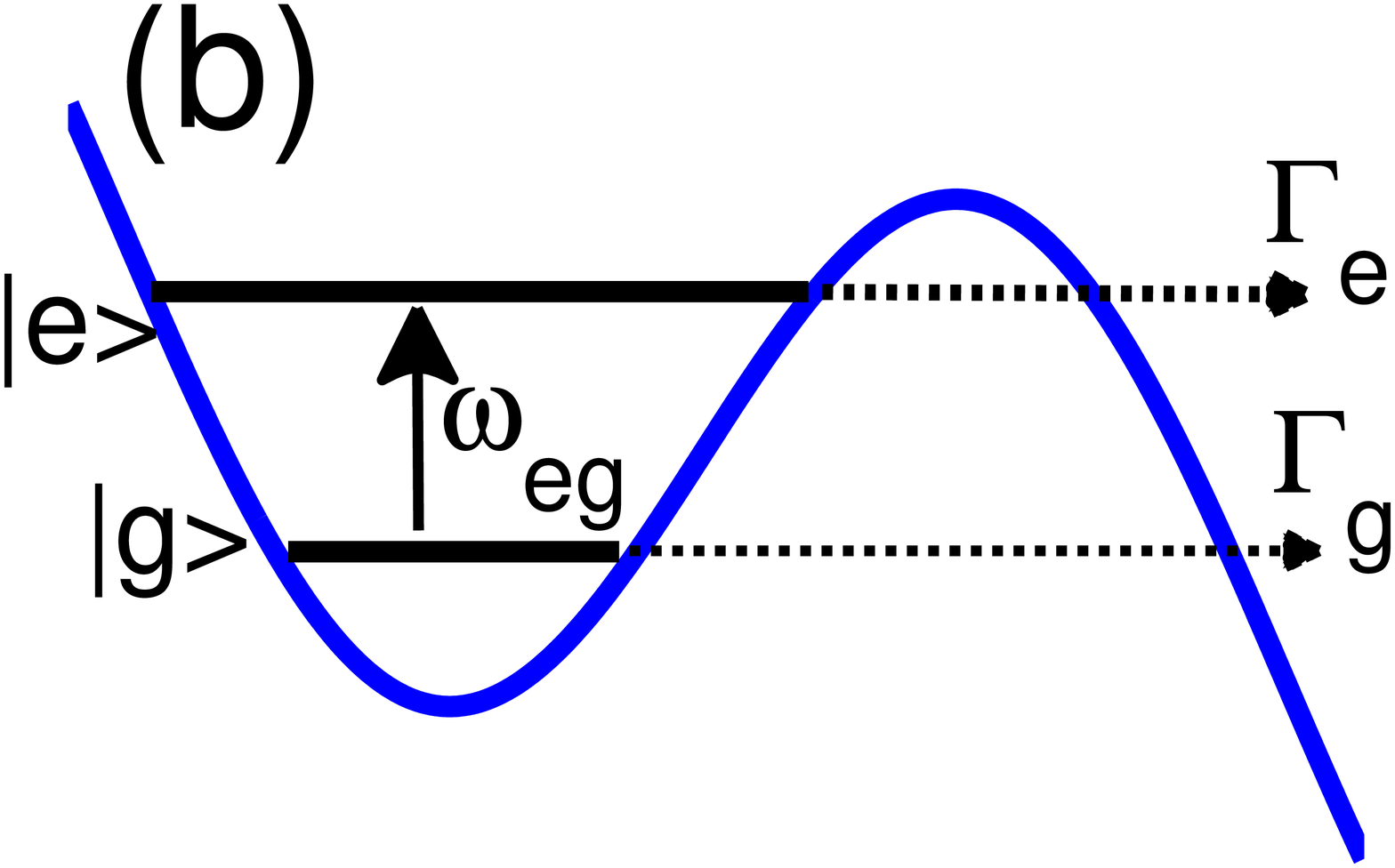}
\caption{(a) Schematic circuit diagram of JJ-based microwave photon detector coupled to a resonator. (b) Potential energy landscape of the detector when bias current is close to the critical current of the JJ . The junction is initialized in the $|g\rangle$ state and upon absorbing an incident photon transitions to the $|e\rangle$ state, which rapidly tunnels to the continuum.}
\label{cubicPot1}
\end{figure}
\begin{figure}
\includegraphics[width=1.0\columnwidth]{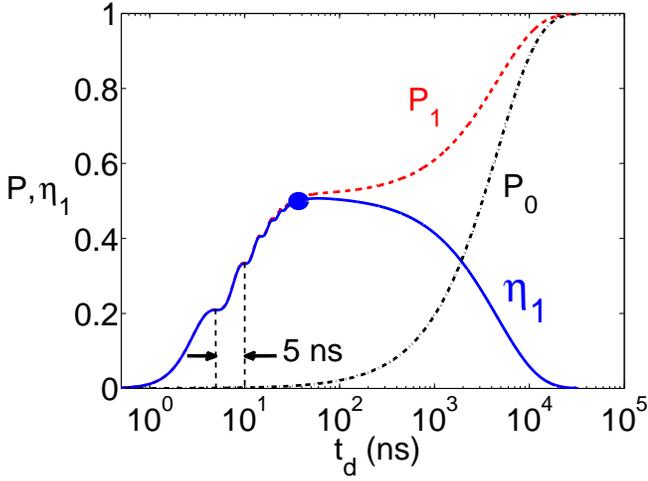}
\caption{(Color online) Switching probability $P_1$ ($P_0$) $vs.$ photon detection time for initial states $|1, g\rangle$ ($|0, g\rangle$). Parameters used in this plot are:  junction $T_1$ = 10\,ns, barrier height $\Delta U/\hbar\omega_p$ = 2, vacuum Rabi frequency $\Omega/2\pi$ =  200\,MHz, and cavity decay time $1/\kappa = 1 \, \mu$s.  Here the detuning between the cavity and the junction $\Delta$ = 0. The solid blue curve is the quantum efficiency $\eta_1 \equiv P_1 - P_0$ (see text). The maximum quantum efficiency of the detector is 50\% for an optimal detection time of 45\,ns (the optimal point is marked by the filled blue circle).}
\label{f1}
\end{figure}

In this work, we theoretically determine the quantum efficiency of a microwave photon detector based on a current-biased JJ. Previous analysis~\cite{Osberg09, Chen11,Peropadre11} of this system was focused on a wave-packet formulation of the photon field in a transmission line coupled to the detector. Here we study the probability of photon detection by a JJ coupled to a microwave cavity mode that is loaded with a fixed number of photons~\cite{You07}. We solve the equation of motion for the density matrix of the cavity-JJ system to obtain detector efficiency for different values of operation time, current bias, and relaxation time of the junction. Our results indicate that a JJ with decay time around 10\,ns can detect a single microwave photon in the cavity with an efficiency greater than 80\%, for readily achievable circuit parameters. We also find that the detector efficiency increases significantly with increasing energy relaxation time $T_1$ of the junction, suggesting that a highly efficient single microwave photon detector is attainable for moderate improvements in junction quality.

\section{Josephson-junction based photon detector}

The circuit diagram of the JJ-cavity system is shown in Fig.~\ref{cubicPot1}(a).  The JJ is biased with a current $I$ close to the critical current $I_0$. The junction Hamiltonian can be written in terms of  the charge operator $\hat{Q}$ and the operator $\hat{\delta}$ of the phase difference across the JJ~\cite{Devoret84}:
\begin{equation}
	\label{eq:HamJ}
	\hat{H}_{JJ} = \frac{\hat{Q}^2}{2C} + U(\hat{\delta}) \,, \quad
	U(\hat{\delta}) = -\frac{I_0 \Phi_0}{2 \pi} \left(\cos \hat{\delta} - \frac{I}{I_0} \hat{\delta}\right).
\end{equation}
Here $C$ is  the junction capacitance and $\Phi_0 = h/2e$ is the magnetic flux quantum. For $I \lesssim I_0$, the potential energy landscape $U(\delta)$ takes on a ``tilted washboard" shape, with a few discrete energy levels in shallow minima separated from the continuum by a barrier, see Fig.~\ref{cubicPot1}(b). We truncate the junction Hamiltonian to the ground $|g\rangle$ and 
first excited states $|e\rangle$ and obtain the following Hamiltonian for the JJ:
\begin{align}
	\hat{H}_{JJ} =  \hbar \omega_{eg}\hat \Pi_e\;,
\end{align}
where $\hat \Pi_e = |e\rangle\langle e|$ is the projection operator to the excited state and
$\omega_{eg} = (\varepsilon_{e}-\varepsilon_g)/\hbar$.

The coupling of the cavity with the JJ is modeled by the Jaynes-Cummings (JC) Hamiltonian~\cite{Jaynes63}:
\begin{equation}
	\label{eq:HamJC}
	\hat{H}_{JC} =  \hbar \omega_r \left(\hat{a}^\dag \hat{a} +\frac{1}{2}\right) + \hbar \omega_{eg} \hat{\Pi}_e +  \frac{\hbar\Omega}{2} (\hat{a}^\dag \hat{\sigma}_{-} + \hat{a} \hat{\sigma}_{+} )\, ,
\end{equation}
where $\omega_r$ is the cavity resonance frequency, $\Omega$ is the vacuum Rabi frequency, and $\hat{a}^{\dag} (\hat{a})$, $\hat{\sigma}_{+} (\hat{\sigma}_{-} )$ are the creation (annihilation) operators of the cavity and the junction, respectively.

The time evolution of the density matrix $\hat\rho(t)$ of the cavity-JJ system coupled to its environment is governed by the following equation:
\begin{align}
	\label{eq:masterEq}
   	 \frac{d\hat{\rho}(t)}{dt} = \frac{1}{i\hbar} \Big[\hat{H}_{JC},\, \hat{\rho}(t)\Big] +  \hat{\mathcal{L}}_{\gamma}[\hat{\rho}(t)] + \hat{\mathcal{L}}_{\kappa}[\hat{\rho}		(t)] + 	\hat{\mathcal{L}}_T[\hat{\rho}(t)] \, ,
\end{align}
where $\hat{\mathcal{L}}_{\gamma}[\hat{\rho}(t)]$ and $\hat{\mathcal{L}}_{\kappa}[\hat{\rho}(t)]$ are superoperators that capture damping in the JJ and the cavity at low temperatures $T\ll \hbar\omega_{eg}$, $\hbar \omega_r$~\cite{Puri86}:
\begin{subequations}
	\begin{align}
	\hat{\mathcal{L}}_{\kappa}[\hat{\rho}(t)] = \kappa
	\left( \hat{a} \hat{\rho} \hat{a}^\dag- \frac{1}{2} \{\hat{a}^\dag \hat{a}, \hat{\rho}\}\right)\,,
	\\
	\hat{\mathcal{L}}_{\gamma}[\hat{\rho}(t)] = \gamma
	\left(\hat{\sigma}_{-} \hat{\rho} \hat{\sigma}_{+} - \frac{1}{2} \{\hat{\sigma}_{+} \hat{\sigma}_{-}, \hat{\rho}\}\right)\,.
\end{align}
\end{subequations}

To account for switching of the ground and the first excited states of the JJ to the voltage state, we introduce the tunneling superoperator $\mathcal{L}_T[\hat{\rho}(t)]$~\cite{, Gurvitz96, Gurvitz98, Ping11, 2Ping11}:
\begin{widetext}
\begin{align}
& \mathcal{\hat{L}}_T[\hat{\rho}(t)]  = -
\left(
\begin{array}{cc}
\displaystyle
\Gamma_e\rho_{ee}^{nn}+\frac{\sqrt{\Gamma_e\Gamma_g}}{2}(\rho_{eg}^{nm}+\rho_{ge}^{nm}) &  
\displaystyle
\frac{\Gamma_e+\Gamma_g}{2}\rho_{eg}^{nm}+\frac{\sqrt{\Gamma_e\Gamma_g}}{2}(\rho_{ee}^{nm}+\rho_{gg}^{nm})\\ 
\displaystyle
\frac{\Gamma_e+\Gamma_g}{2}\rho_{ge}^{nm}+\frac{\sqrt{\Gamma_e\Gamma_g}}{2}(\rho_{ee}^{nm}+\rho_{gg}^{nm}) &
\displaystyle \Gamma_g\rho_{gg}^{nn}+\frac{\sqrt{\Gamma_e\Gamma_g}}{2}(\rho_{eg}^{nm}+\rho_{ge}^{nm})
\end{array} 
\right) \, ,
\end{align}
\end{widetext}
where $\Gamma_{e,g}$ are the tunneling rates from the ground ($|g\rangle$) and  first excited ($|e\rangle$) states of the junction.  If we approximate the potential in Eq.~\eqref{eq:HamJ} by a cubic potential, then the tunneling rate of the ground and the first excited states of the cubic potential can be computed by WKB approximation:
\be
\Gamma_ j = \omega_p/2\pi [432 \Delta U/\hbar \omega_p]^{j+1/2}/\pi^{j/2} \exp[-36 \Delta U/5 \hbar \omega_p]\,, \nonumber 
\ee
where $\Gamma_{j=0}\equiv \Gamma_g$ and $\Gamma_{j=1}\equiv \Gamma_e$ represent the tunneling rates from the $|g\rangle$ and $|e\rangle$ states of the JJ, respectively. The ratio $\Gamma_e/\Gamma_g \approx 250 \Delta U/\hbar \omega_p$. Here $\Delta U = 4 I_0 \Phi_0/3\sqrt{2}\pi(1- I/I_0)^{3/2}$ is the barrier height and $\omega_p = 2^{1/4}\,\sqrt{2\pi I_0/ C \Phi_0} (1-I/I_0)^{1/4}$ is the plasma frequency of the cubic potential. The junction frequency $\omega_{eg}$ is related to the plasma frequency by $\omega_{eg} \simeq \omega_p(1-5\hbar\omega_p/36\Delta U)$. The tunneling rate of the first excited state of the junction is then given by $\Gamma_e \approx 500\,\Gamma_g = 7.3\times 10^7$\,s$^{-1}$ for $\Delta U/\hbar\omega_p \approx 2$.

\section{Quantum efficiency}

The system is originally prepared in a pure state $\hat\rho_n(0)=\ket{n,g}\bra{n,g}$ with $n$ photons in the cavity and the junction in the ground state $|g\rangle$. We assume $n$ photons are loaded into the cavity in a manner similar to that described by Hofheinz $et$ $al.$~\cite{Hofheinz08}, with loading rate faster than the interaction rate between the JJ and the cavity. With this assumption, the detector efficiency is unaffected by how photons are loaded into the cavity. We numerically solve the above equation for the time evolution of the density matrix to compute the occupation probabilities of the cavity and junction states.  The probability that the JJ has switched to the voltage state at time $t$ is given by
\begin{equation}
P_n(t)=1 -  \Tr\,[\hat{\rho}_n(t)].
\end{equation}
\begin{figure}
\includegraphics[width=1.0\columnwidth]{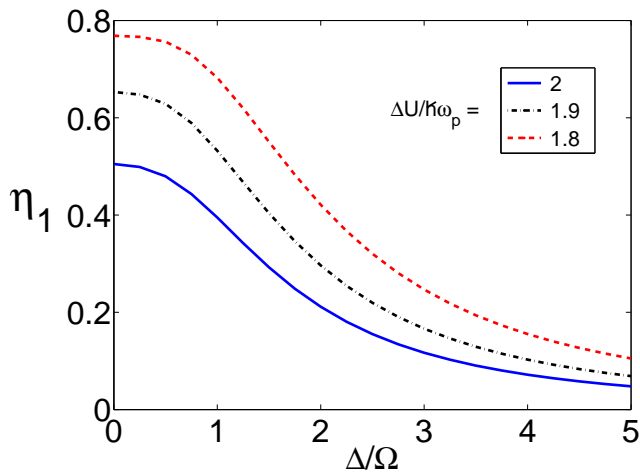}
\caption{(Color online) Quantum efficiency $\eta_1$ $vs.$ detuning $\Delta/\Omega$ for optimal detection time obtained at $\Delta= 0$ for various bias points: $\Delta U/\hbar \omega_p = 2$ (solid blue),  1.9 (dashed-dot black), and 1.8 (dashed red). The bandwidths of the detector are 1.6$\Omega$ (solid blue), 2$\Omega$ (solid black) and 2.3$\Omega$ (solid red), respectively. The remaining parameters are as in Fig.~\ref{f1}.}
\label{detune}
\end{figure}

\begin{figure}[t]
 \includegraphics[width=1.0\columnwidth]{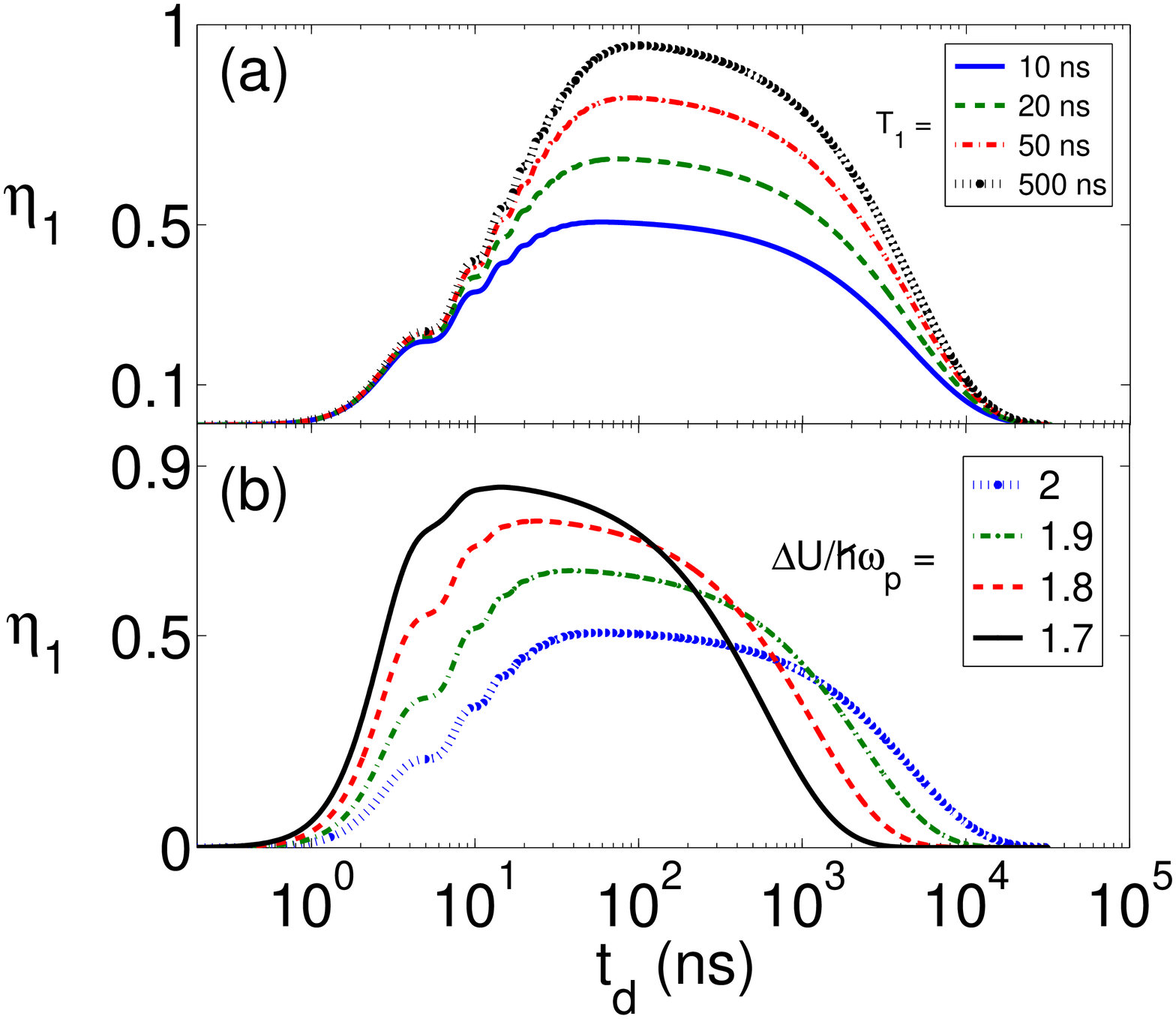}
\caption{(Color online)(a) Quantum efficiency $\eta_1$ $vs.$ photon detection time for $\Delta U/\hbar\omega_p$ = 2 and for various decay times of the junction: $T_1$ = 10\,ns (solid blue), $T_1$ = 20\,ns (dashed green), $T_1$ =  50\,ns (dash-dotted red), and $T_1$ = 500\,ns (dotted black). For $T_1 = $ 500 ns, the maximum quantum efficiency is 94\%. (b) Quantum efficiency $\eta_1$ $vs.$ photon detection time for junction $T_1$ = 10\,ns for various bias points of the junction: $\Delta U/\hbar\omega_p$ = 2 (dotted blue), 1.9 (dash-dotted green), 1.8 (dashed red) and 1.7 (solid black). The maximum quantum efficiency is 84\% for $\Delta U/\hbar\omega_p$ = 1.7. The remaining parameters are as in Fig.~\ref{f1}. }
\label{f2}
\end{figure}

\begin{figure}[t]
 \includegraphics[width=1.0\columnwidth]{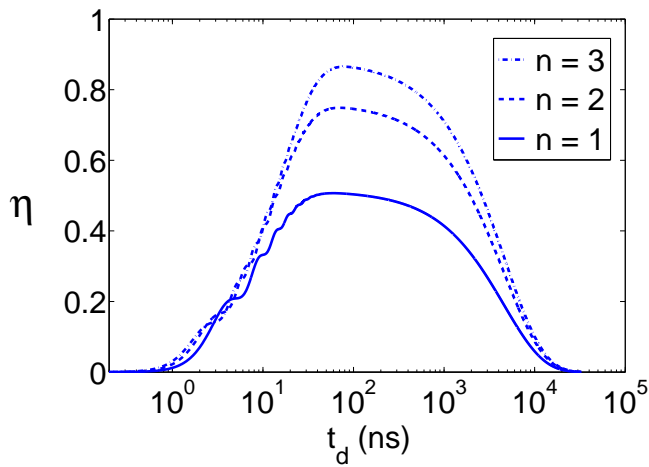}
 \caption{(Color online) Efficiency $\eta_n$ to detect a photon $vs.$ detection time for $n$ = 1 (solid blue), 2 (dashed blue), and 3 (dash-dotted blue) photons in the cavity. The rest of the parameters are as in Fig.~\ref{f1}. For three photons in the cavity, the efficiency to detect a photon is 85\%.}
 \label{f3}
 \end{figure}
We consider the following set of parameters for the JJ-cavity system, typical of those realized in experiments~\cite{Chen11}: JJ frequency $\omega_{eg}/2\pi$ = 4.8\,GHz, junction decay rate $\gamma$ = $10^8$\,s$^{-1}$, cavity decay rate  $\kappa$ = $10^6$\,s$^{-1}$, and vacuum Rabi frequency $\Omega/2\pi$ = 200\,MHz. 

In Fig.~\ref{f1}, we plot switching probabilities $P_1(t), P_0(t)$ of the junction for initial states $|1, g \rangle$ (solid red) and $|0, g \rangle$ (solid black), respectively. In this simulation, the parameters we consider are $\Delta U/\hbar\omega_p$ = 2 and we set the detuning between the cavity and the JJ to zero, i.e., $\Delta \equiv \omega_r - \omega_{eg} = 0$. The switching probability $P_1$  in Fig.~\ref{f1} features steps whose periodic occurrence is a manifestation of Rabi oscillations of the JJ with period $2\pi/\Omega = 5$\,ns.  This result is consistent with the picture that switching of the JJ halts momentarily when the junction returns to the ground state in course of the Rabi oscillations.

Next, we discuss the presence of the wide plateau of $P_1$ in Fig.~\ref{f1}. The occurrence of this plateau can be understood from the fact that switching of the junction is briefly frozen when the junction relaxes to the ground state due to dissipation. The JJ then switches to the voltage state after time $\sim 1/\Gamma_g$, the characteristic time scale for switching of the junction in the case of zero photons. The height of this plateau can be estimated as $\Gamma_e/\big(\Gamma_e + \gamma\big) \approx$ 0.5, which agrees with the numerical result in Fig.~\ref{f1}.

In order to determine the quantum efficiency of the detector, we must properly treat dark counts due to quantum tunneling from the $\ket{0,g}$ state in the absence of photon absorption. The quantum efficiency $\eta_1$ of the detector is defined as the difference between the switching probabilities for an initial state with one photon $P_1(t)$, and for an initial state with no photons $P_0(t)$: $\eta_1 \equiv P_1(t) - P_0(t)$. The quantum efficiency is shown in Fig.~\ref{f1} by the solid blue curve. For our choice of parameters $\Gamma_e \simeq \gamma$, the detector has maximum efficiency of about 50\% for the optimal detection time $t_d$ around 45\,ns. 

Next, we demonstrate that the bandwidth of the Josephson microwave photon detector is broadened due to the finite lifetime of the junction excited state. Here, we vary the frequency $\omega_r$ of the cavity and compute the quantum efficiency of the detector for the optimal detection time $t_d$ obtained at zero detuning $\Delta$ = 0. The detector bandwidth is then given by the detuning at which the quantum efficiency of the detector is reduced to half the efficiency obtained at zero detuning.
For a dissipation--free junction, the detector bandwidth is approximately given by the vacuum Rabi frequency. However, in the presence of dissipation and tunneling, the first excited state of the junction is broadened by $\sim \gamma + \Gamma_e$.
This broadening of the energy level roughly accounts for the increased bandwidth of the detector.  We find that bandwidths are factors of 1.6,  2 and 2.3 larger than the vacuum Rabi frequency for bias points $\Delta U/\hbar \omega_p = 2, 1.9$ and $1.8$, respectively, as shown in Fig.~\ref{detune}. As we lower the ratio $\Delta U/\hbar \omega_p$, the tunneling rate $\Gamma_e$ of the first excited state of the junction increases. This in turn causes further broadening of the junction excited state thereby increasing the bandwidth of the detector.

Next, we analyze the effect of dissipation and bias point on the efficiency of the detector. In Fig.~\ref{f2}(a), we plot the quantum efficiency of the detector for different values of the junction relaxation time $T_1$ from 10 ns to 500 ns, keeping all other parameters the same as in Fig.~\ref{f1}. We find that the quantum efficiency increases with increasing junction relaxation time $T_1$ and reaches 94\% for $T_1= 500$\,ns and for a detection time around 95 ns.

The change in bias current $I$ of the junction modifies the ratio of barrier height $\Delta U$ to the junction plasma frequency $\omega_p$.  Taking different values of this ratio, we compute the efficiency of the detector at fixed relaxation time $T_1 = 10$ ns; the results are shown in Fig.~\ref{f2} (b). Upon decreasing the ratio $\Delta U/\hbar \omega_p$, the potential well becomes shallower, leading to enhanced tunneling out of the first excited state and increased efficiency of the detector. Our simulation results indicate that a significant improvement in detector efficiency is achieved when the tunneling rate exceeds the dissipation rate of the junction. We find that for the bias point $\Delta U/\hbar \omega_p$ = 1.7, the efficiency of the detector is about 84\% for a detection time around 9 ns.

Finally, we analyze the efficiency of the detector to detect single photons when the cavity is loaded with $n > 1$ photons. Generalizing the previous case of a single photon in the cavity, the efficiency to detect a single microwave photon in a cavity loaded with $n$ photons is given by $\eta_n=P_n(t) - P_0(t)$.  In Fig.~\ref{f3}, we plot the efficiency at fixed bias point $\Delta U/\hbar \omega_p =2$ and $T_1 = 10$ ns for different numbers of photons in the cavity: $n$ = 1 (solid blue), 2 (dashed blue), and 3 (dash-dotted blue). We find that detection efficiency increases with the increasing number of photons in the cavity and reaches 85\% for three photons in the resonator. This result is consistent with previous studies~\cite{Osberg09, Chen11} that reported a higher detection efficiency, for the same parameters as above, when a continuous flux of photons was incident on the detector. For the case of a single photon in the cavity, the detector returns to the ground state after the photon is absorbed by the environment, and no further excitation of the junction is possible. However, when multiple photons are present in the cavity, other photons are available to induce excitation if the junction relaxes, thereby increasing the probability of photon detection. We note that for multiple photons in the cavity, the measured efficiency $\eta$ can also be used to estimate the average number of photons in the cavity.

 \section {Discussion and Conclusions}
In conclusion, we have presented a model to determine the quantum efficiency of a microwave photon detector based on a current-biased JJ. 
We have demonstrated that the efficiency to detect a single photon loaded in a cavity has maximal value $\Gamma_e/(\Gamma_e+\gamma)$. We have also determined that the bandwidth of the detector is characterized by the sum of the vacuum Rabi frequency and the broadening of the first excited state of the JJ due to tunneling and relaxation processes.  Our simulations indicate that for currently used  JJ photon detectors, the quantum efficiency is about  50\% for the bias point $\Delta U/\hbar \omega_p$ = 2 and about 85\% for $\Delta U/\hbar\omega_p$ = 1.7.  We have finally investigated the probability to detect a photon in the case of a multiphoton initial resonator state and have found that the detection efficiency quickly approaches 100\% as the initial number of photons increases, consistent with previous studies~\cite{Osberg09, Chen11} of a continuous flux of photons incident on the detector.

Recently, Peropadre $et$ $al.$~\cite{Peropadre11} proposed a phenomenological model for the JJ-based microwave photon detector that fails to address specific microscopic details of experimentally realized detectors~\cite{Chen11}. Specifically: (1) Peropadre $et$ $al.$ treat  tunneling from the excited state of the junction by a non-hermitian term in the junction Hamiltonian; this is not consistent with the standard form of quantum tunneling. (2) These authors do not consider tunneling from the low-energy state of the junction, which is responsible for dark counts of the detector. (3) Finally, their model does not take into account the relatively strong relaxation from the excited to the ground state of the junction.  This relaxation corresponds to a $T_1$ time of order of a few nanoseconds in present devices, which are strongly coupled to a 50 $\Omega$ readout line, and is responsible for a significant suppression of detector efficiency~\cite{Chen11}.  If the relaxation time  were above 500 ns, the efficiency would reach nearly 100\%, see Fig.~4(a).

\acknowledgements

We gratefully acknowledge discussion with Guilhem Ribeill. This work was supported by the ARO under grant W911NF-11-1-0030 and by the NSF under grant DMR-1105178.

\end{document}